\documentclass[aps,twocolumn,prl,superscriptaddress,showpacs,nofootinbib]{revtex4}

\usepackage{graphicx}
\usepackage{ulem}
\usepackage{dcolumn}
\usepackage{ifthen}
\usepackage{natbib}
\usepackage{verbatim}
\usepackage{amsmath}

\providecommand{\e}[1]{\ensuremath{\times 10^{#1}}}

\begin{document}

\title{Parity violation constraints using 2006-2007 QUaD CMB polarization spectra}

\author{E.\,Y.\,S.\,Wu}
\email{e2wu@stanford.edu}
\affiliation{Kavli Institute for Particle Astrophysics and Cosmology and Department of Physics, Stanford University, Stanford, CA 94305, USA.}

\author{P.\,Ade}
\affiliation{School of Physics and Astronomy, Cardiff University, Cardiff CF24 3AA, UK.}

\author{J.\,Bock}
\affiliation{California Institute of Technology, Pasadena, CA 91125, USA.}
\affiliation{Jet Propulsion Laboratory, Pasadena, CA 91109, USA.}

\author{M.\,Bowden}
\affiliation{School of Physics and Astronomy, Cardiff University, Queen's Buildings, The Parade, Cardiff CF24 3AA, UK.}
\affiliation{Kavli Institute for Particle Astrophysics and Cosmology and Department of Physics, Stanford University, Stanford, CA 94305, USA.}

\author{M.\,L.\,Brown}
\affiliation{Cavendish Laboratory, University of Cambridge, Cambridge CB3 OHE, UK.}

\author{G.\,Cahill}
\affiliation{Department of Experimental Physics, National University of Ireland Maynooth, Maynooth, Co. Kildare, Ireland.}

\author{P.\,G.\,Castro} 
\affiliation{CENTRA, Departamento de F\'{\i}sica, Universidade T\'ecnica de Lisboa, 1049-001 Lisboa, Portugal.}  
\affiliation{Institute for Astronomy, University of Edinburgh, Edinburgh EH9 3HJ, UK.}

\author{S.\,Church}
\affiliation{Kavli Institute for Particle Astrophysics and Cosmology and Department of Physics, Stanford University, Stanford, CA 94305, USA.}

\author{T.\,Culverhouse}
\affiliation{Kavli Institute for Cosmological Physics,
  Department of Astronomy \& Astrophysics, Enrico Fermi Institute, University of Chicago,Chicago, IL 60637, USA.}

\author{R.\,B.\,Friedman}
\affiliation{Kavli Institute for Cosmological Physics,
  Department of Astronomy \& Astrophysics, Enrico Fermi Institute, University of Chicago,Chicago, IL 60637, USA.}

\author{K.\,Ganga}
\affiliation{Laboratoire APC/CNRS, B\^atiment Condorcet, 75205 Paris Cedex 13, France.}

\author{W.\,K.\,Gear}
\affiliation{School of Physics and Astronomy, Cardiff University, Cardiff CF24 3AA, UK.}

\author{S.\,Gupta}
\affiliation{School of Physics and Astronomy, Cardiff University, Cardiff CF24 3AA, UK.}

\author{J.\,Hinderks}
\affiliation{Kavli Institute for Particle Astrophysics and Cosmology and Department of Physics, Stanford University, Stanford, CA 94305, USA.}

\author{J.\,Kovac}
\author{A.\,E.\,Lange}
\affiliation{California Institute of Technology, Pasadena, CA 91125, USA.}

\author{E.\,Leitch}
\affiliation{California Institute of Technology, Pasadena, CA 91125, USA.}
\affiliation{Jet Propulsion Laboratory, Pasadena, CA 91109, USA.}

\author{S.\,J.\,Melhuish}
\affiliation{School of Physics and Astronomy, University of Manchester, Manchester M13 9PL, UK.}

\author{Y.\,Memari}
\affiliation{Institute for Astronomy, University of Edinburgh, Edinburgh EH9 3HJ, UK.}

\author{J.\,A.\,Murphy}
\affiliation{Department of Experimental Physics, National University of Ireland Maynooth, Maynooth, Co. Kildare, Ireland.}

\author{A.\,Orlando}
\affiliation{California Institute of Technology, Pasadena, CA 91125, USA.}
\affiliation{School of Physics and Astronomy, Cardiff University, Cardiff CF24 3AA, UK.}

\author{L.\,Piccirillo}
\affiliation{School of Physics and Astronomy, University of Manchester, Manchester M13 9PL, UK.}

\author{C.\,Pryke}
\affiliation{Kavli Institute for Cosmological Physics,
  Department of Astronomy \& Astrophysics, Enrico Fermi Institute, University of Chicago,Chicago, IL 60637, USA.}

\author{N.\,Rajguru}
\altaffiliation{Department of Physics and Astronomy, University College London, Gower Street, London WC1E 6BT, UK.}
\affiliation{School of Physics and Astronomy, Cardiff University, Cardiff CF24 3AA, UK.}

\author{B.\,Rusholme}
\altaffiliation{Infrared Processing and Analysis Center,
  California Institute of Technology, Pasadena, CA 91125, USA.}
\affiliation{Kavli Institute for Particle Astrophysics and Cosmology and Department of Physics, Stanford University, Stanford, CA 94305, USA.}

\author{R.\,Schwarz}
\affiliation{Kavli Institute for Cosmological Physics,
  Department of Astronomy \& Astrophysics, Enrico Fermi Institute, University of Chicago,Chicago, IL 60637, USA.}

\author{C.\,O'\,Sullivan}
\affiliation{Department of Experimental Physics, National University of Ireland Maynooth, Maynooth, Co. Kildare, Ireland.}

\author{A.\,N.\,Taylor}
\affiliation{Institute for Astronomy, University of Edinburgh, Edinburgh EH9 3HJ, UK.}

\author{K.\,L.\,Thompson}
\affiliation{Kavli Institute for Particle Astrophysics and Cosmology and Department of Physics, Stanford University, Stanford, CA 94305, USA.}

\author{A.\,H.\,Turner}
\affiliation{School of Physics and Astronomy, Cardiff University, Cardiff CF24 3AA, UK.}

\author{M.\,Zemcov}
\affiliation{Jet Propulsion Laboratory, Pasadena, CA 91109, USA.}
\affiliation{California Institute of Technology, Pasadena, CA 91125, USA.}
\affiliation{School of Physics and Astronomy, Cardiff University, Cardiff CF24 3AA, UK.}

\collaboration{The QUaD Collaboration}
\noaffiliation

\begin{abstract}
We constrain parity-violating interactions to the surface of last
scattering using spectra from the QUaD experiment's second and third
seasons of observations by searching for a possible systematic
rotation of the polarization directions of CMB photons. We measure the
rotation angle due to such a possible ``cosmological birefringence''
to be $0.55^\circ \pm 0.82^\circ$ (random) $\pm 0.5^\circ$
(systematic) using QUaD's 100 and 150~GHz TB and EB spectra over the
multipole range $200<\ell<2000$, consistent with null, and constrain
Lorentz violating interactions to $< 2\e{-43}$ GeV (68$\%$ confidence
limit).  This is the best constraint to date on electrodynamic parity
violation on cosmological scales.
\end{abstract}



\maketitle

\section{Background}

Cosmic Microwave Background (CMB) polarization measurements at
multipoles of $\ell > 20$ are unaffected by reionization and are an
effective means to probe for cosmological scale electrodynamic parity
violation to the surface of last scattering.  Using the CMB is
particularly attractive because of the long path length to the surface
of last scattering, the well-understood physics of the primordial
universe that generated the CMB photons, and two cross-spectra, the
temperature-curl (TB) and gradient-curl (EB) cross-correlations, that
should be null in a parity-conserving universe \citep{Lue, Lepora,
  1997PhRvD..55.7368K, 1997PhRvL..78.2058K, 1997PhRvD..55.1830Z}. As
the effect should be frequency independent, measurements of the CMB
at multiple frequencies can distinguish it from other EB correlation inducing
effects like Faraday rotation from magnetic fields in the
intergalactic medium \citep{Faraday, Gardner, 1997PhRvD..56.7493S}.

The known parity violation in the weak force is sufficient motivation
for investigating electrodynamic parity violation, but it has been
shown that parity-violating interactions are a potential solution to
the problem of baryon number asymmetry because they can be a signature
of CPT (charge-parity-time) violation in an expanding universe
\citep{Feng05}.  

The effect arises by adding a Cherns-Simons term to
the normal electrodynamic Lagrangian, violating Lorentz, P and CPT
symmetries \citep{Carroll, Xia_postWMAP5}:

\begin{equation}
\mathcal{L} = -\frac{1}{4} F_{\mu\nu}F^{\mu\nu} + p_{\mu}A_{\nu}\tilde{F}^{\mu\nu}
\end{equation}

Here $F^{\mu\nu}$ denotes the field tensor, $\tilde{F}^{\mu\nu}$ is
its dual, $p_{\mu}$ is an external vector, and $A_{\nu}$ the 4-vector
potential. Non-zero time or space components of $p_{\mu}$ induce a
rotation of the polarization direction of each photon as it propagates
from the surface of last scattering.  This is equivalent to a local
rotation of the Stokes parameters, Q and U, in the polarization maps
made by CMB experiments, inducing gradient (E) to curl (B) mode mixing
and therefore EB correlation.  Lorentz violation can also be tested
with these models \citep{Carroll, Kostelecky}. In addition, models of
quintessence can be probed by examining the EB and TB spectra for
non-zero power \citep{2006PhRvL..97p1303L}.

QUaD was a 100 and 150 GHz bolometric polarimeter that made deep
observations of the CMB from the South Pole during the austral winters
of 2005 through 2007.  A recent analysis of the second and third
seasons of data from QUaD shows a series of acoustic peaks in the EE
auto-spectra over the multipole range $200<\ell<2000$ consistent with
the $\Lambda$-CDM model of the universe~\citep{Pryke}.  This dataset
offers the strongest constraining power to date on cosmological scale
parity-violating interactions.  The QUaD collaboration maintains two
code independent, but nearly algorithmically identical data analysis
pipelines for the purposes of consistency checking. The results
presented here use the 100 and 150~GHz spectra from the ``alternative
pipeline'' described in section 6.8 of \citet{Pryke} for reasons of
computational convenience, derived using a modified version of the
MASTER CMB analysis method \citep{Hivon}.

\section{Analysis}
Assuming that the CMB is a Gaussian random field, the entirety of its
statistical properties can be described by the auto- and
cross-correlation power spectra:

\begin{equation}
C_\ell^{XY} = \frac{1}{2\ell + 1}\sum_m a_{\ell m}^{X*} a_{\ell m}^{Y}
\end{equation}

\noindent where the $a_{\ell m}$ are the coefficients of the spherical
harmonic decomposition of the temperature or polarization maps. $X$
and $Y$ here denote $T$, $E$ or $B$ for the respective maps of
temperature, gradient-polarization and curl-polarization modes.

Normally the $C_\ell^{TB}$ and $C_\ell^{EB}$ are expected to be null
because the spherical harmonic eigenfunctions $Y_{\ell m}^{T}$ and
$Y_{\ell m}^{E}$ have parity $(-1)^\ell$ and $Y_{\ell m}^{B}$ has
parity $(-1)^{\ell + 1}$.  Assuming that there is a parity-violating
effect in the electrodynamics equations that prefers one polarization
to another over cosmological scales, let us denote the average
preferred rotation of the polarization direction of a photon from the
surface of last scattering as it heads towards us as
$\Delta\alpha$. This corresponds to a rotation of the polarization
directions in the maps~\citep{Lue, Feng05} inducing E to B mixing, and
therefore EB cross correlation.  Likewise, since there is already TE
cross correlation, TB cross correlation is also induced.  Following
\citet{Komatsu}, we assume that cosmological BB modes are zero to
simplify the equations and maximize the likelihood of a detection:

\begin{eqnarray}
C_\ell^{TE,obs} &=& C_\ell^{TE} \cos(2\Delta\alpha) \\
C_\ell^{TB,obs} &=& C_\ell^{TE} \sin(2\Delta\alpha) \\
C_\ell^{EE,obs} &=& C_\ell^{EE} \cos^2 (2\Delta\alpha)  \\
C_\ell^{BB,obs} &=& C_\ell^{EE} \sin^2 (2\Delta\alpha) \\
C_\ell^{EB,obs} &=& \frac{1}{2} (C_\ell^{EE}) \sin(4\Delta\alpha)
\end{eqnarray}

For the purposes of plotting and analysis, we can derive a
theory-independent $\chi^2$ statistic to combine the first two and the
last three equations separately to obtain an estimate of
$\Delta\alpha$, utilizing constraining power from across our 23
reported bandpowers. First, we assume $\ell(\ell + 1)C_\ell^{XX,obs}$
is constant within a bandpower and define the quantities below for
each bandpower:

\begin{eqnarray}
D_{TB,\ell} &=& C_\ell^{TB,obs}\cos \left(2 \Delta\alpha \right) - C_\ell^{TE,obs} \sin \left(2 \Delta\alpha \right)\\
D_{EB,\ell} &=& C_\ell^{EB,obs} \\ \nonumber && -\frac{1}{2} (C_\ell^{BB,obs} + C_\ell^{EE,obs}) \sin \left(4 \Delta\alpha \right)
\end{eqnarray}

We can then minimize $\chi^2(\Delta\alpha)$ for the TB and EB
combinations separately to estimate $\Delta\alpha$ \footnote{It is
  also possible to estimate $\Delta\alpha$ by measuring the quantities
  $\frac{2 C_\ell^{EB,obs}}{(C_\ell^{EE,obs} + C_\ell^{BB,obs})}$ and
  $\frac{C_\ell^{TB,obs}}{\sqrt{(C_\ell^{TE,obs})^2 +
      (C_\ell^{TB,obs})^2}}$ on a per-bandpower basis, combining them
  using the covariances as measured from simulations, and then
  applying inverse trigonometric functions. However, this is biased
  in the presence of noise. We thank an anonymous referee for suggesting our current method.}:

\begin{eqnarray}
\chi^2 (\Delta\alpha) &=& \sum_{\ell\ell^{`}}D_{TB,\ell} M_{\ell\ell^{'}}^{-1} D_{TB,\ell^{'}}\\
\chi^2 (\Delta\alpha) &=& \sum_{\ell\ell^{`}}D_{EB,\ell} M_{\ell\ell^{'}}^{-1} D_{EB,\ell^{'}}
\end{eqnarray}

\noindent We empirically measure the covariance matrix, $M_{\ell\ell^{'}}$, of
the bandpowers in each spectrum $D_{EB,\ell}$ and $D_{TB,\ell}$ from a
set of simulated bandpowers combining realizations of $\Lambda$-CDM
cosmology temperature and polarization fields for the signal component
and accurate realizations of QUaD's instrumental noise. Our method
utilizes a set of 496 signal and noise Monte Carlo simulations from
the analysis pipeline of QUaD.  \citet{Pryke} demonstrates
the robustness of QUaD's simulation method against a variety of
systematics tests.

\newcolumntype{d}{D{,}{\pm}{8.13}}
\newcolumntype{e}{D{,}{\pm}{7.7}}
\begin{table*}[hbt]
\centering
\caption{Column 1: $\Delta\alpha$ measurements from QUaD, including
  random and systematic errors. Column 2: Bias and standard errors on
  mean sampled from 496 signal-only simulations. Column 3: Column 2
  subtracted from Column 1. Column 4: Scatter of signal-only
  simulations, indicating sample variance. Column 5: Fraction of
  signal+noise simulations where $\|\Delta\alpha\|$ exceeds that of
  data.}  
\begin{tabular}{c | d e d c c}
\hline\hline
 & \multicolumn{1}{c}{$\Delta\alpha$} & \multicolumn{1}{c}{systematic} & \multicolumn{1}{c}{bias-corrected $\Delta\alpha$} & signal-only & $\%$ simulations \\
Spectrum & \multicolumn{1}{c}{(random and sys. errors)} & \multicolumn{1}{c}{bias} & \multicolumn{1}{c}{(random and sys. errors)} & simulation scatter & exceeding \\
\hline
150 GHz EB       &  0.76^\circ,0.92^\circ\pm0.5^\circ &  0.003^\circ,0.003^\circ &  0.76^\circ,0.92^\circ\pm0.5^\circ & $0.08^\circ$ & 41.3\% \\
150 GHz TB       &  1.19^\circ,3.26^\circ\pm0.5^\circ &  0.025^\circ,0.017^\circ &  1.16^\circ,3.26^\circ\pm0.5^\circ & $0.37^\circ$ & 71.5\% \\
100 GHz EB       & -3.74^\circ,2.22^\circ\pm0.5^\circ &  0.011^\circ,0.004^\circ & -3.75^\circ,2.22^\circ\pm0.5^\circ & $0.10^\circ$ & 8.87\% \\
100 GHz TB       &  3.72^\circ,5.69^\circ\pm0.5^\circ &  0.073^\circ,0.022^\circ &  3.65^\circ,5.69^\circ\pm0.5^\circ & $0.50^\circ$ & 52.2\% \\
150 GHz combined &  0.85^\circ,0.94^\circ\pm0.5^\circ &  0.015^\circ,0.003^\circ &  0.83^\circ,0.94^\circ\pm0.5^\circ & $0.07^\circ$ & 35.8\% \\
100 GHz combined & -1.86^\circ,2.24^\circ\pm0.5^\circ &  0.031^\circ,0.005^\circ & -1.89^\circ,2.24^\circ\pm0.5^\circ & $0.11^\circ$ & 38.7\% \\
100/150 combined &  0.56^\circ,0.82^\circ\pm0.5^\circ &  0.011^\circ,0.004^\circ &  0.55^\circ,0.82^\circ\pm0.5^\circ & $0.08^\circ$ & 49.6\% \\
\hline
\end{tabular}
\label{table:da}
\end{table*}

\begin{figure}[htb]
\centering
\caption{$\Delta\alpha$ measured from QUaD 150 GHz TB and EB spectra;
  histogram of simulations and red line for data. Histogram does not
  account for systematic error. Dotted line indicates total
  uncertainty assuming a Gaussian $0.5^\circ$ systematic error.}
\includegraphics[width=1.0\columnwidth]{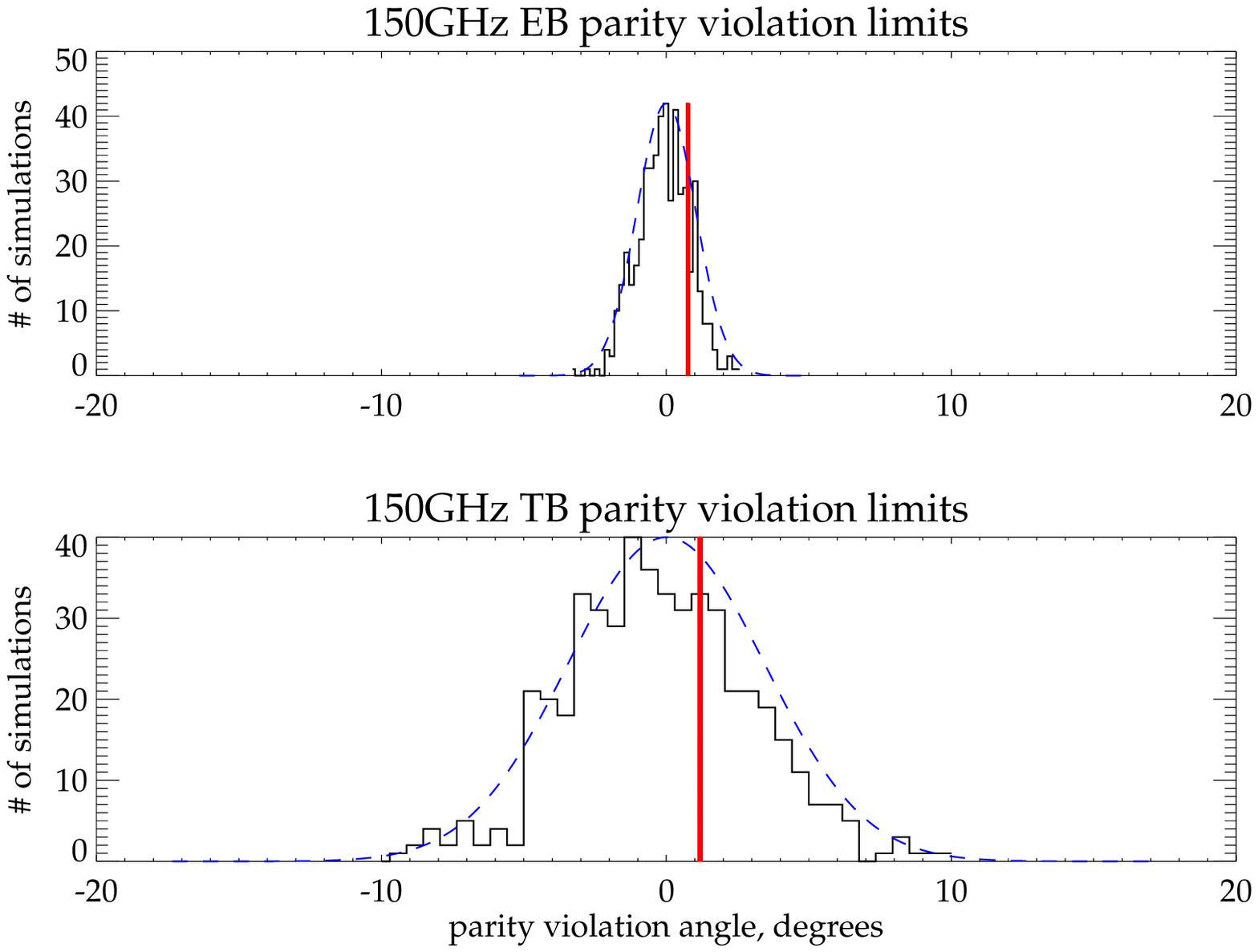}
\label{150GHz_limits}
\end{figure}

\begin{figure}[htb]
\centering
\caption{150 GHz $\Delta\alpha$ per bandpower derived from the EB spectrum.
  Note that in practice these 
  points are combined before the final transformation to
  $\Delta\alpha$ --- the purpose of this plot is to give a visual
  representation of the relative uncertainties across the bandpowers. }

\includegraphics[width=1.0\columnwidth]{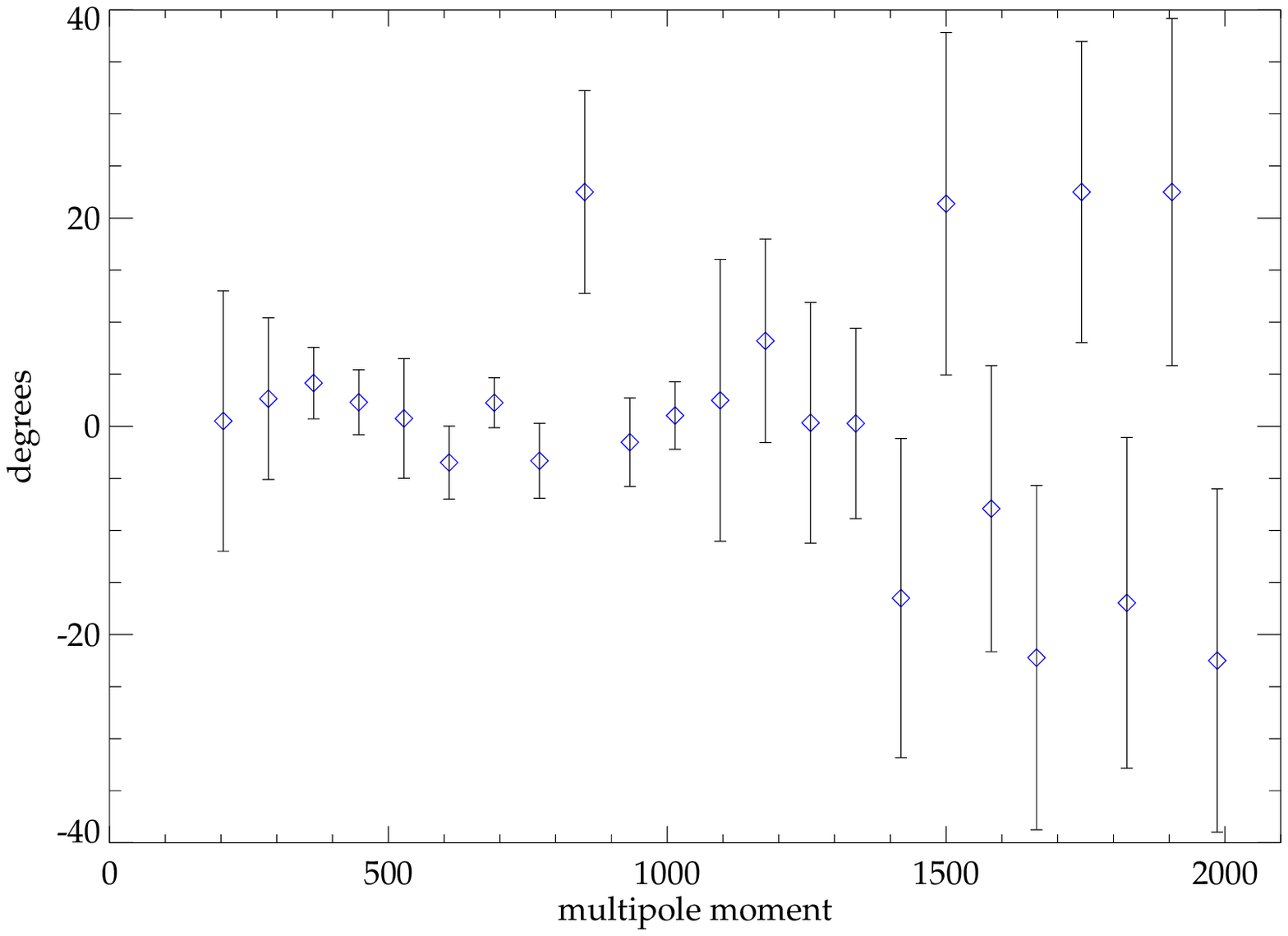}
\label{150GHzEB}
\end{figure}

Figure \ref{150GHz_limits} shows the results of this combination for
the data (red line) and simulations (histogram) for 150GHz in both EB
and TB. Overplotted is the total uncertainty assuming that the
simulations reflect a normal distribution and that the systematic
error is $0.5^\circ$. It is clear that the observed data can easily be
drawn from the set of simulations in which no parity-violating
interactions have been included; we therefore conclude that there is
no detection.

To obtain a visual representation of a ``$\Delta\alpha$ spectrum,'' We
can also estimate the best fit for $\Delta\alpha$ on a per-bandpower
basis by minimizing:

\begin{eqnarray}
\chi^2_{\ell} (\Delta\alpha) &=& \sum_{\ell^{`}}D_{TB,\ell} M_{\ell\ell^{'}}^{-1} D_{TB,\ell^{'}}\\
\chi^2_{\ell} (\Delta\alpha) &=& \sum_{\ell^{`}}D_{EB,\ell} M_{\ell\ell^{'}}^{-1} D_{EB,\ell^{'}}
\end{eqnarray}

\noindent The $\Delta\alpha$ spectrum using the EB, BB and EE spectra
for 150GHz is shown in figure \ref{150GHzEB}.

\section{Current Limits and QUaD Results}
Komatsu et. al. \cite{Komatsu} report their limits from the WMAP 5
year high-$\ell$ data as $\Delta\alpha = -1.2^\circ \pm 2.2^\circ$.
Other authors have found weak evidence for parity violation by
combining the WMAP 5 year data and data from the BOOMERanG balloon
experiment, reporting $\Delta\alpha = -2.6^\circ \pm 1.9^\circ$
\citep{Xia_postWMAP5}.  Carroll et. al. \citet{Carroll} derived
constraints on $\Delta\alpha$  10 high-redshift radio galaxies in
1990, yielding $\Delta\alpha = -0.6^\circ \pm 1.5^\circ$. The best
single redshift number, for 3C9 at $z = 2.012$, is $\Delta\alpha =
2^\circ \pm 3^\circ$.

QUaD's results broken down by individual spectrum and frequency, as
well as combined within and between frequencies, are shown in Table
\ref{table:da}. Reported errors are $68.2\%$ confidence limits as
determined by the distribution of signal and noise simulations. 150
GHz EB alone is significantly more constraining than any current
result.  At no frequency, nor in any spectrum, is there a significant
detection. We also present values for $\Delta\alpha$ where the
systematic bias induced by a combination of timestream filtering and
the slightly different, non-aligned, and elliptical nature of the
beams of two orthogonally aligned polarization sensitive detectors
within a single feedhorn leading to temperature to polarization
leakage has been quantified by signal-only simulations. This effect is
discussed in further detail in \citet{Hinderks}. Note that in all
frequencies and spectra this bias is an order of magnitude smaller
than our random and systematic errors.  After combination the EB
spectra dominate the analysis and there is virtually no bias. These
results are consistent with a constraint on isotropic
Lorentz-violating interactions of $k^{(3)}_{(V)00} < 2\e{-43}$ GeV
\citep{Kostelecky}.

\section{Systematic effects and checks}

The primary systematics concern is that there might be a systematic
rotation of the true detector sensitivity angles, producing a false
signal totally degenerate with that of parity violation; for example,
a $-3^\circ$ systematic misalignment and a $\Delta\alpha=-3^\circ$
true parity violation signal would produce identical results. We have
measured the overall rotation angle of our instrument using two
methods. The first measures the polarization sensitivity angle of each
bolometer using a near field polarization source. The second
constrains the absolute angle of the focal plane by examining the
measured offsets of the beams of each detector from the telescope
pointing direction on an astronomical source. These two methods agree
nearly exactly indicating that any systematic rotation of the
bolometers within the focal plane structure is negligible.  Given that
there is no physical or mechanical reason to suspect such a rotation,
and the uncertainties of the measurements, we conservatively assign a
systematic uncertainty on the absolute rotation angle of the
instrument of $0.5^\circ$ and quote this value in the abstract and in
Table~\ref{table:da}.

We have reanalyzed the entire dataset after inserting an artificial 2
degree local polarization rotation only in the data maps, resulting in
a 2 degree shift after deriving $\Delta\alpha$ identically to the
procedure above, validating the analysis pipeline.

A secondary concern is random scatter in the assumed detector
angles. This is a different effect than a systematic rotation of all
of the detectors.  The Monte Carlo simulation pipeline includes the
injection of a degree of uncertainty about the true orientation of
each polarization sensitive bolometer (PSB) into every simulation
commensurate with the uncertainty of the measurements described in
\citet{Hinderks}. Thus, when constructing a ``fake focal plane'' for
signal-only simulations of a given CMB realization, we assign every
bolometer a random deviation from its presumed angle at
reconstruction, drawn from a Gaussian distribution with $\sigma =
1^\circ$. We also assume that the polarization grids are sensitive to
the orthogonal polarization direction at a level of $7 \pm 3\%$ and
include this effect in the simulation pipeline. Signal-only
simulations with and without these effects included show that their
contribution to the final uncertainty is small.

As detailed in Table~\ref{table:da}, our analysis of signal-only
simulations reveal a small bias in the recovered $\Delta\alpha$
values. In order to isolate the source of this bias, we have performed
additional sets of signal-only simulations, including in isolation the
effects of filtering, mis-aligned beams, uncertainties in detector
alignment and cross-polar leakage. The results from these tests
confirm that a combination of timestream filtering and beam
mis-alignment is the source of the bias. Note that, although small
compared to our noise-driven errors, our results do include a
correction for the bias.

\section{Conclusions}
We have presented the strongest constraints on parity violation to
date. Assuming that there are no cosmological-scale parity violating
interactions, we have also demonstrated that it is possible to
understand the cumulative effects of detector misalignment
uncertainties in polarization sensitive bolometer-based instruments to
under $1^\circ$ through a combination of analysis of primary CMB
polarization data and lab measurements. This is of potential interest
with respect to analysis of data from the High Frequency Instrument of
the upcoming Planck Satellite.

We thank the substantial contributions of an anonymous referee for
improving our method.  QUaD is funded by the National Science
Foundation in the USA, through grants ANT-0637420, ANT-0739729,
ANT-0638615, ANT-0638352, ANT-0739413, AST-0096778, ANT-0338138,
ANT-0338335 and ANT-0338238, by the UK Science and Technology
Facilities Council (STFC) and its predecessor the Particle Physics and
Astronomy Research Council (PPARC), and by the Science Foundation
Ireland.  EYW acknowledges receipt of an NDSEG fellowship.  JRH
acknowledges the support of an NSF Graduate Research Fellowship, a
Stanford Graduate Fellowship and a NASA Postdoctoral Fellowship.  PGC
is funded by the {\it Funda\c{c}\~ao para a Ci\^encia e a
  Tecnologia}. CP and JEC acknowledge partial support from the Kavli
Institute for Cosmological Physics from the grant NSF PHY-0114422.  MZ
acknowledges the support of a NASA Postdoctoral Fellowship.

\bibliographystyle{apsrev}
\newcommand{\mnras}{Monthly Notices of the Royal Astronomical Society}
\newcommand{\apjl}{Astrophysical Journal Letters}

\end{document}